\documentclass[aps,pra,twocolumn,showpacs,superscriptaddress,floatfix,longbibliography]{revtex4-1}

\usepackage[utf8]{inputenc} 
\usepackage[T1]{fontenc}
\usepackage{amssymb,amsmath}
\usepackage{graphicx}
\usepackage{braket}
\usepackage{siunitx}
\usepackage{xcolor}
\usepackage{physics}
\usepackage{hyperref}

\usepackage{xcolor}
\usepackage[normalem]{ulem}

\def\kr{k_{\mathrm R}}                            			
\def\Er{E_{\mathrm R}}                            			
\def\Rb87{^{87}\mathrm{Rb}}                             
\def\K40{^{40}\mathrm{K}}                    		    
\def\Ud{U_{\rm D}}
\def\Udd{U}
\def\Ad{{\mathbf A}_{\rm D}}
\def\Add{{\mathbf A}}

\def\Bdd{{\mathbf B}}
\def\psid{\psi_{\rm D}}
\def\psib{\psi_{\rm B}}
\def\psie{\psi_{\rm E}}

\def\ex{\boldsymbol{e}_x}  
\def\ey{\boldsymbol{e}_y}  
\def\ez{\boldsymbol{e}_z}

\begin{document}

\title{Interference induced anisotropy in a two-dimensional dark state optical lattice}

\author{E.~Gvozdiovas}
\affiliation{Institute of Theoretical Physics and Astronomy, Vilnius University, Saulėtekio Ave. 3, LT-10257 Vilnius, Lithuania}

\author{I.~B.~Spielman}
\email{ian.spielman@nist.gov}
\homepage{http://ultracold.jqi.umd.edu}
\affiliation{Joint Quantum Institute, University of Maryland, College Park, Maryland 20742-4111, USA}
\affiliation{National Institute of Standards and Technology, Gaithersburg, Maryland 20899, USA}

\author{G.~Juzeliūnas}
\email{gediminas.juzeliunas@tfai.vu.lt}
\affiliation{Institute of Theoretical Physics and Astronomy, Vilnius University, Saulėtekio Ave. 3, LT-10257 Vilnius, Lithuania}

\date{\today}

\begin{abstract}
    We describe a two-dimensional optical lattice for ultracold atoms with spatial features below the diffraction limit created by a bichromatic optical standing wave.
    At every point in space these fields couple the internal atomic states in a three-level Lambda coupling configuration.
    Adiabatically following the local wavefunction of the resulting dark state yields a spatially uniform Born-Oppenheimer potential augmented by
    geometric scalar and vector potentials appearing due to spatially rapid changes of the wavefunction.
    Depending on system parameters, we find that the geometric scalar potential can interpolate from a 2D analogue of the Kronig-Penney lattice, to an array of tubes with a zig-zag shaped barrier.
    The geometric vector potential induces a spatially periodic effective magnetic field (the Berry's curvature) that can be tuned to cause destructive interference between neighboring tubes, thereby decoupling them at a critical point in parameter space.
    We numerically investigate the energy spectrum including decay from the excited state, and find that the adiabatic approximation is sound for strong coupling strengths, leading to negligible loss in the dark state manifold.
    Furthermore, the spectrum is well-described by a non-Hermitian tight binding model with on-site losses, and hopping characterized by both loss and, surprisingly, gain.
\end{abstract}

\maketitle


Realizing long-lived strongly correlated quantum matter with ultracold atoms in optical lattices is an ongoing challenge.
Despite the now decades old realization of the superfluid to Mott insulator transition in 1D, 2D and 3D \cite{Greiner2002,Stoferle2004,Spielman2007}, there has been little progress in realizing strongly correlated systems such as fractional quantum Hall states.
In both cases, interactions are enhanced by reducing the contribution of the kinetic energy: inhibiting tunneling in a deep optical lattice in the case of a Mott insulator, or quenching the kinetic energy with a magnetic field in the case of fractional quantum Hall states.
The former case simply localizes particles to lattice sites, producing an uncorrelated insulator. 
Here we describe a new technique for creating nearly flat bands, even in the presence of strong tunneling, using Aharonov-Bohm like quantum interference between sites.

We consider a 2D extension to existing 1D dark state optical lattices studied theoretically \cite{Zoller2016, Jendrzejewski2016, Zubairy2020, Kubala2021, Gvozdiovas2021}, and realized experimentally \cite{Wang2018, Tsui2020}, enabling interference phenomena that are not possible in 1D.
As indicated in Figs.~\ref{fig:setup}(a,b), our lattice is created from a pair of orthogonal standing waves and a transverse running wave coupling three internal atomic internal states in a Lambda configuration.
The local dark state of this scheme has zero energy and no excited state contribution.
Atomic motion introduces geometric scalar and vector potentials \cite{Dum1996,Juzeliunas2005,Goldman2014}, as well as non-adiabatic mixing to the excited state.
The geometric potentials are maximal at the nodes of the optical standing wave where the atomic dark state changes rapidly.
For atoms adiabatically following the dark state in 1D, this gave rise to a Kronig-Penney like lattice with barriers far narrower than the optical wavelength \cite{Zoller2016, Jendrzejewski2016,Wang2018}.
For specific parameters we find the natural 2D analog of this lattice consisting of square tiles spaced by narrow barriers.
However, generically the geometric  scalar and vector potentials---the latter quantified by the Berry curvature---can form a lattice of Dirac $\delta$-function like needles, or can take on a serpentine appearance, creating an array of undulating tubes.
Unexpectedly, we observe that these tubes abruptly decouple at critical points in parameter space where Aharonov-Bohm interference from the geometric vector potential inhibits tunneling.
This produces nearly completely flat bands transverse to the tubes with barriers that would otherwise allow substantial tunneling.
The band flattening is analogous to the formation of dispersionless Landau levels with the application of a uniform magnetic field.

Figure~\ref{fig:setup}(b) schematically illustrates our proposed experimental geometry.
A laser beam traveling along $\ez$ drives one arm of a $\Lambda$-scheme that is then completed by a second arm consisting of four mutually interfering laser beams in the $\ex$-$\ey$ plane.
This geometry adds two additional degrees of freedom as compared to the existing 1D dark-state lattices: the relative intensity intensity between the in-plane lasers as well as their relative phase (controlled by displacing a retro-reflection mirror).
Changing the relative intensity converts linear barriers in to serpentine ones.
Tuning the relative phase morphs the lattice from a 2D Kronig-Penney like lattice of linear barriers to one with needle-like potential maxima.
In addition, the phase difference breaks time reversal symmetry, introducing a non-zero Berry curvature.

Our manuscript is organized as follows.
We introduce the basic formulation of our 2D $\Lambda$-lattice in Sec.~\ref{sec:formulation}, and identify the associated symmetries in Sec.~\ref{sec:symmetries}.
Section~\ref{sec:results} describes our numerical method, and presents our main results.
Lastly, in Sec.~\ref{sec:outlook} we conclude with a discussion and outlook.

\section{Formulation}\label{sec:formulation}

\subsection{Hamiltonian for the 2D Lambda scheme }

\begin{figure}[t]
\centering
  \includegraphics[scale=1,keepaspectratio]{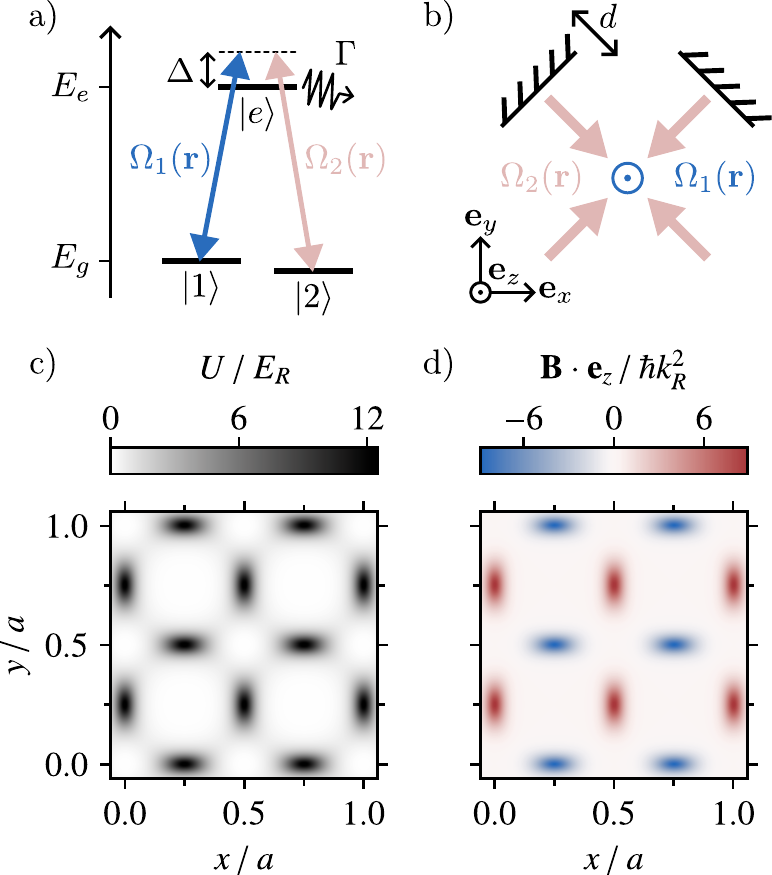}
  \caption{a), Lambda($\Lambda$) coupling scheme.
Two atomic ground states $\protect\ket1$ and $\protect\ket2$ are laser-coupled with detuning $\Delta$ to an excited state $\protect\ket e$ (with spontaneous decay rate $\Gamma$)
with strengths $\Omega_{1,2}(\boldsymbol{r})$.
b), Schematic. $\Omega_{1}(\boldsymbol{r})$ is a plane wave traveling along $\ez$ and $\Omega_{2}(\boldsymbol{r})$ consists of two orthogonal standing waves created by the pictured interfering beams.
A movable mirror imparts a controllable phase shift on the retro-reflected beam traveling along $\ex-\ey$.
c), Geometric scalar $\Udd$ and d), Berry curvature $\Bdd$ for motion in the dark state computed for $\epsilon = 0.4$, $\epsilon_c = 1$ and $\chi = \pi / 4$. }
  \label{fig:setup}
\end{figure}

We consider ultracold atoms subject to a 2D atom-light interaction 
\begin{equation}
\frac{\hat{V}(\boldsymbol{r})}{\hbar}=\left(-\Delta-\frac{i}{2}\Gamma\right)\ket e\bra e+\sum_{j=1}^{2}\left[\;  \frac{\Omega_{j}(\boldsymbol{r})}{2}\ket e\bra j+ {\rm H.c.} \right]\,,\label{eq:V(r)-Operator}
\end{equation}
describing the $\Lambda$-type coupling scheme \cite{Dum1996,Juzeliunas2005,Goldman2014} shown in Fig.~\ref{fig:setup}(a).
Here the atomic ground states $\ket 1$ and $\ket 2$ are coupled with strength $\Omega_{j}(\boldsymbol{r})$ to the excited state $\ket e$ with detuning $\Delta$~\footnote{Here we adopt the rather unnatural ``spectroscopy'' convention where positive $\Delta$ corresponds to negative energy difference for transitions to the excited state.}.
The excited state has a spontaneous decay rate $\Gamma$.
Altogether this gives the atomic Hamiltonian
\begin{equation}
\hat{H}=\frac{\hat{\mathbf{p}}^{2}}{2M}+\hat{V}(\boldsymbol{r}) \label{eq:H}
\end{equation}
in terms of the position $\boldsymbol{r} = (x,y)$, momentum $\hat{\mathbf{p}} = - i \hbar \boldsymbol{\nabla}$, and atomic mass $M$.
The above Hamiltonian is non-Hermitian due to the imaginary contribution $i \Gamma / 2$ arising from time-irreversible decay from the excited state $\ket{e}$ in Eq.~\eqref{eq:V(r)-Operator}.   
In Sec.~\ref{sec_energy_dispersions} we numerically demonstrate that losses due to $i \Gamma / 2$ are minimal in the so-called dark state and thus Hermitian dynamics are maintained. Additionally, even the Hermitian contribution to \eqref{eq:V(r)-Operator} can break time reversal symmetry when any of the $\Omega_j$ coefficients are complex.

The Hamiltonian acts in the space of state-vectors
\begin{equation}
\ket{\psi\left(\boldsymbol{r}\right)} = \sum_{j=1,2,e} \psi_{j}\left(\boldsymbol{r}\right) \ket{j} \, , \label{eq:|psi>-expansion}
\end{equation}
containing the atomic internal states $\ket{j}$ and the associated wave-functions $\psi_{j}\left(\boldsymbol{r}\right)$ for atomic center of mass motion.
The corresponding full abstract state vector would be given by $\ket{\psi} = \int d \boldsymbol{r} \ket{\psi\left(\boldsymbol{r}\right)} \otimes \ket{\boldsymbol{r}} $, with $\ket{\psi\left(\boldsymbol{r}\right)} = \braket{\boldsymbol{r}}{\psi}$.

\subsection{New basis with dark state}

We now re-express $\ket{\psi\left(\boldsymbol{r}\right)}$ in a basis containing a long-lived dark state in which geometric potentials with sub-wavelength features can emerge.
A dark state is a (generally position-dependent) superposition of atomic ground states for which $\hat{V}(\boldsymbol{r})\ket{D}=0$.
Therefore, in such a basis, $\hat{V}(\boldsymbol{r})$ contributes no potential energy, no coupling terms, and no spontaneous decay for $\ket{D}$. 
This allows the two coupling arms in Fig.~\ref{fig:setup}(a) to be driven on resonance with $\ket{e}$ without loss from $\ket{D}$.

Here we consider orthogonal dark
\begin{equation}
\ket{D(\boldsymbol{r})} = \frac{1}{\Omega} \left[ \; \Omega_{2}(\boldsymbol{r}) \ket{1}  - \Omega_{1}(\boldsymbol{r}) \ket{2} \; \right]
\label{eq:D}
\end{equation}
and bright
\begin{equation}
\ket{B(\boldsymbol{r})} = \frac{1}{\Omega} \left[ \;  \Omega_{1}(\boldsymbol{r}) \ket{1}  + \Omega_{2}(\boldsymbol{r}) \ket{2} \; \right]
\label{eq:B}
\end{equation}
state superpositions, where $\Omega = \sqrt{ \left| \Omega_{1} \right|^2 + \left| \Omega_{2} \right|^2 }$ is an averaged coupling strength.
Unlike $\ket{D(\boldsymbol{r})}$, the bright state couples to the excited state $\ket{e}$.

In the basis of dark, bright and excited states, the state vector is \begin{equation}
\ket{\psi\left(\boldsymbol{r}\right)} = \psid\left(\boldsymbol{r}\right) \ket{D(\boldsymbol{r})} + \psib\left(\boldsymbol{r}\right) \ket{B(\boldsymbol{r})} + \psie\left(\boldsymbol{r}\right) \ket{e}\,,\label{eq:|psi>-expansion-dark, bright, excited}
\end{equation}
where $\psid\left(\boldsymbol{r}\right)$, $\psib\left(\boldsymbol{r}\right)$
and $\psie\left(\boldsymbol{r}\right)$ are wave-functions for
the atomic center of motion in the corresponding internal states.

The atom-light coupling operator $\hat{V}(\boldsymbol{r})$ is diagonalized by the trio of dressed states $\ket{D}$ and $\ket{\pm}$, where $\ket{\pm}$  are superpositions of $\ket{B}$ and $\ket{e}$ only.
When these states depend on position, they are not eigenstates of the full Hamiltonian $\hat H$ due to the kinetic energy $\hat{\mathbf{p}}^{2} / (2 M)$; this leads to geometric potentials for the projected dynamics in each dressed state~\cite{Goldman2014,Mead1992}.
In the present case, geometric potentials are introduced by the spatially varying coupling strengths $\Omega_{1,2}(\boldsymbol{r})$.

\subsection{Effective potentials for adiabatic dark state}
\label{sec_effective_adiabatic}

When the total Rabi frequency $\Omega$ at every point in space greatly exceeds the characteristic energy of the atomic center of mass motion, the atoms will adiabatically
follow their initial dressed state with negligible transitions to the other dressed states.
For dark-state atoms, the state vector \eqref{eq:|psi>-expansion-dark, bright, excited} can be approximated as 
\begin{equation}
|\psi\left(\boldsymbol{r}\right)\rangle\approx\psid\left(\boldsymbol{r}\right) \ket{D(\boldsymbol{r})}\,.
\label{eq:psi_D}
\end{equation}
The validity of this approximation for 1D dark state lattices has been extensively studied~\cite{Zoller2016, Zubairy2020, Jendrzejewski2016}.
We correspondingly arrive at the 2D adiabatic Hamiltonian in the dark state manifold \cite{Juzeliunas2005,Goldman2014} 
\begin{equation}
\hat{H}_{\rm D}=\frac{1}{2M}(-i\hbar\boldsymbol{\nabla}-\Ad)^{2}+\Ud\,,
\label{eq:Schroed_D}
\end{equation}
where $\Ud$ and $\Ad$ are the geometric
scalar and vector potentials.
Because our focus is on the dark state manifold, we suppress the subscript ${\rm D}$ in what follows.
The scalar potential
\begin{equation}
\Ud \equiv \Udd(\boldsymbol{r})=\frac{\hbar^{2}}{2M}\frac{(\boldsymbol{\nabla}\xi^{\ast})\cdot(\boldsymbol{\nabla}\xi)}{\left(1+|\xi|^{2}\right)^{2}} \,
\label{eq:U_D_general}
\end{equation}
is plotted in Fig.~\ref{fig:setup}(c) and Fig.~\ref{Fig_2}. Here we introduce $\xi(\boldsymbol{r}) \equiv \Omega_{2}(\boldsymbol{r}) / \Omega_{1}(\boldsymbol{r})$, the complex valued ratio of coupling strengths in Eq.~\eqref{eq:D}.
The geometric vector potential
\begin{equation}
\Ad \equiv \Add(\boldsymbol{r})=i\hbar\frac{\xi^{\ast}\boldsymbol{\nabla}\xi-\xi\boldsymbol{\nabla}\xi^{\ast}}{2\left(1+|\xi|^{2}\right)} \,
\label{eq:A_D_general}
\end{equation}
is non-zero only when $\xi(\boldsymbol{r})$ has an imaginary component.
Lastly, the geometric magnetic field shown in Fig.~\ref{fig:setup}(d) is the curl of the vector potential
\begin{equation}
\Bdd(\boldsymbol{r})=\boldsymbol{\nabla}\times\Add=i\hbar\frac{(\boldsymbol{\nabla}\xi^{\ast})\times(\boldsymbol{\nabla}\xi)}{\left(1+|\xi|^{2}\right)^{2}} \, .
\label{eq:B_D_general}
\end{equation}
We confirm the adiabatic assumption for our 2D lattice in Sec.~\ref{sec_energy_dispersions} by comparing numerical results from the full non-Hermitian Hamiltonian \eqref{eq:H} to the adiabatic approximation (Fig.~\ref{Fig_comparison}).

\subsection{Sub-wavelength effective potentials}
\label{sec_effective_subwavelength}

\begin{figure*}[t]
\centering
  \includegraphics[scale=1,keepaspectratio]{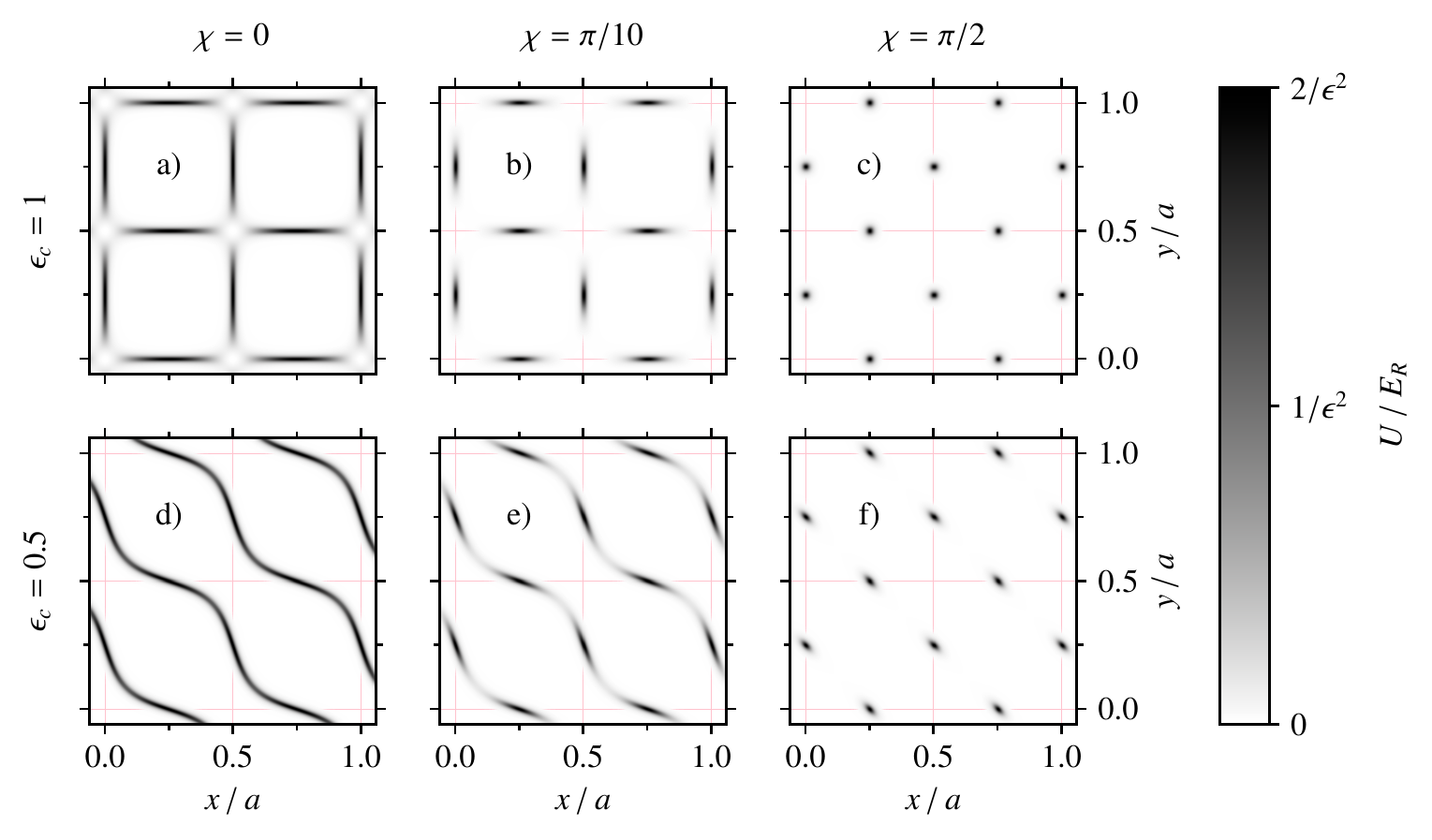}
  \caption{Geometric scalar potential $\Udd$ for various $\chi$ and $\epsilon_c$ with $\epsilon=0.1$.}
  \label{Fig_2}
\end{figure*}

We now describe a configuration of couplings $\Omega_{1,2}(\boldsymbol{r})$ shown in Fig.~\ref{fig:setup}(b) which yield rapid changes in the dark state wavefunction \eqref{eq:D}, parameterized by $\boldsymbol{\nabla}\xi$ [entering Eqs.~\eqref{eq:U_D_general}--\eqref{eq:B_D_general}], resulting in 2D geometric potentials with features below the optical diffraction limit.
The first laser field $\Omega_{1}(\boldsymbol{r}) = \Omega_p \, \exp(i \kr z)$ coupling $\ket{1}$ to $\ket{e}$ is a plane wave traveling along $\ez$ with amplitude $\Omega_{p}$ and wavevector $\kr=2\pi/a$.
We consider atoms that are tightly confined in the $z=0$ plane, such that $\exp(i \kr z) \approx 1$.
The second coupling field $\Omega_{2}(\boldsymbol{r})$ results from a crossed pair of standing waves with amplitudes $\Omega^{(\pm)}_{c}$ in the $\ex$-$\ey$ plane \cite{Hemmerich1991,Hemmerich1992,Juz-Spielm2012NJP}; a controllable path-length-difference $d$ for the retro-reflected field along $(\ex-\ey)/\sqrt{2}$ introduces a phase $\chi=d/a$ which breaks time-reversible symmetry and allows for non-zero Berry curvature.

After making the rotating wave approximation (RWA) and phase shifting $\ket{2}$ by $e^{-i \chi / 2}$, the Rabi frequencies of the coupling fields are
\begin{align}
& \Omega_{1}(\boldsymbol{r}) = \Omega_p \, , \label{Rabi_frequencies_0} \\
& \Omega_{2}(\boldsymbol{r})= \sum_\pm \pm \Omega^{(\pm)}_{c} e^{\mp i\chi/2} \cos(\kr x \pm \kr y)\, . \nonumber
\end{align}
Inserting \eqref{Rabi_frequencies_0} into \eqref{eq:U_D_general}--\eqref{eq:B_D_general}, the explicit forms of the geometric potentials for this choice of coupling fields are
\begin{align}
\frac{\Udd (\boldsymbol{r})}{\Er} &= \frac{ \beta^{2}_{+}+ \epsilon^2_c \beta^{2}_{-}}{\alpha^2} 2\epsilon^{2} (1+\epsilon_c^2) \, ,
\nonumber\\
\frac{\Add (\boldsymbol{r})}{\hbar \kr} &= \frac{ \sin{(2\kr y)} \mathbf{e}_x + \sin{(2\kr x) \mathbf{e}_y } }{\alpha} \epsilon_c \sin{\chi} \, ,
\label{eq:A_D} \\
\frac{B_{z} (\boldsymbol{r})}{\hbar \kr^2} &= \frac{ \cos{(2 \kr x)} - \cos{(2 \kr y)}   }{\alpha^2} 2 \epsilon^{2} (1+\epsilon_c^2) \epsilon_c \sin{\chi} \, ,
\nonumber
\end{align}
where $\Er=\hbar^{2}\kr^{2}/(2M)$ is the single photon recoil energy, and $\Bdd(\boldsymbol{r}) = B_{z}(\boldsymbol{r}) \, \ez$ implying that the magnetic field is orthogonal to the $\ex$-$\ey$ plane. 
Here we defined a factor
\begin{align*}
\alpha{(\boldsymbol{r})} &= \epsilon^2 (1 + \epsilon_c^2) + \eta^2_{+} + \epsilon_c^2 \eta^2_{-} - 2 \epsilon_c \eta_+ \eta_- \cos{\chi} \,
\end{align*}
present in \eqref{eq:A_D}, and
\begin{align*}
\eta_{\pm}(\boldsymbol{r}) &= \cos{(\kr x \pm \kr y)} \, , & {\rm} && \beta_{\pm}(\boldsymbol{r}) &= \sin{(\kr x \pm \kr y)} \, ,
\end{align*}
%
%
as well as the ratios of the laser field amplitudes
\begin{align}
\epsilon &= \frac{\Omega_p}{\sqrt{ \Omega_c^{(+)2} + \Omega_c^{(-)2} } }& \text{and} && \epsilon_c &= \frac{\Omega^{(-)}_{c} }{ \Omega^{(+)}_{c}}. \label{eq:epsilon}
\end{align}
The ratios $\epsilon$ and $\epsilon_c$ determine how rapidly the internal structure of the dark state changes near the zeros of $\Omega_2(\boldsymbol{r})$, allowing control of both the height and spatial extent of the effective potentials in Eqs.~\eqref{eq:A_D}.

The scalar potential $\Udd$ is plotted Fig.~\ref{Fig_2} for different values of $\chi$ and $\epsilon_c$ ($B_{z}$, not shown, is graphically very similar to $\Udd$).
The peak values of the scalar and magnetic fields
\begin{align*}
U_{\text{max}} &= \frac{2 \Er}{\epsilon^2} & {\text{and}\rm} && B_{\text{max}}(\boldsymbol{r}) &= \frac{4 \hbar \kr^2 \epsilon_c \sin{\chi}}{\epsilon^2 (1+\epsilon_c^2)} \,
\end{align*}
are proportional to $2/\epsilon^{2}$, and thus diverge as $\epsilon \rightarrow 0$.
Additionally, $B_{\text{max}}$ depends on $\epsilon_c$ and $\chi$, reaching a maximum value with $\epsilon_c = 1$ and $\chi=\pi/2$.
We also numerically computed the full width at half maximum of the maxima of $\Udd$ and $\Bdd$ along their thinnest direction (as seen in Fig.~\ref{Fig_2}(d), this direction has no particular association with $\ex$ or $\ey$).
We find that when $\epsilon\ll1$, i.e., $\Omega_2\gg\Omega_1$, the lattice has tall sub-wavelength barriers that can be further tuned by adjusting $\epsilon_c$ and $\chi$.

The ratio $\epsilon_c$ determines the degree of serpentine bending in the geometric potential.
This leads to an effective 1D to 2D transition; for example (with $\chi=0$), the potential transitions from a brick-like structure (with holes at the crossing points) at $\epsilon_c = 1$ [Fig.~\ref{Fig_2}(a)], to an an array of modulated walls [$\epsilon_c = 0.5$ in Fig.~\ref{Fig_2}(d)], finally arriving at straight 1D walls [$\epsilon_c = 0$].
Additionally, the lattices shown in Fig.~\ref{Fig_2}(d)--(f) can be rotated by 90 degrees by replacing $\epsilon_c \rightarrow 1/\epsilon_c$.

When $\chi=\pi/2$ and $\epsilon_c = 1$ [Fig.~\ref{Fig_2}(c)], the magnetic field and scalar potential reduce to a 2D array of needle-like peaks.
Moreover, with $\epsilon \rightarrow 0$ the 2D integral of $\Udd$ over the peak converges to $\Er a^2 / 2\pi = \hbar^2 \pi / M$.
As such, even for $\epsilon \rightarrow 0$ the surface integral of the scalar potential does not diverge, leading to a 2D array of Dirac $\delta$ function potentials---a 2D Dirac comb---with strength $\Er a^2 / 2\pi$.
The same applies to the magnetic field $\Bdd$ with strength $2 \pi \hbar$.

Just as in the 1D case, intensity imbalances between the different arms of the $\Omega_2({\bf r})$ field ultimately limit the minimum width of the barriers~\cite{Wang2018}.
Since our primary focus is on interference effects rather than minimizing the barrier widths, this is not a significant consideration in this work.

In the next Section we consider the symmetries of the atom-light coupling which impose requirements on the eigensolutions of both the full and dark state Hamiltonians.
These symmetries will be later utilized in the numerical treatment to unfold the energy bands.

\section{Symmetries of the Hamiltonian}
\label{sec:symmetries}

Including the couplings $\Omega_{1,2}$ in Eq.~\eqref{Rabi_frequencies_0}, the full Hamiltonian \eqref{eq:H} is invariant with respect
to spatial shifts along $\ex$ and $\ey$ by the lattice constant $a$, i.e.,  $\hat{H}(x+a,y) = \hat{H}(x,y+a) = \hat{H}(x,y)$, so that
\begin{equation}
\left[\hat{H},\exp\left(  - \frac{ i \, \mathbf{a}_l \cdot \hat{\mathbf{p}}}{\hbar}   \right)\right]=0\,,\quad\mathrm{with}\quad l=1,2\,, \label{eq_commutator_a}
\end{equation}
with elementary unit vectors
\begin{align}
\mathbf{a}_{1} &= a\, \ex, & {\rm and} && \mathbf{a}_{2} &=a\, \ey\,.
\end{align}

Interestingly, the dark state geometric potentials are symmetric with regards to translations by $a/2$, as evident in Figs.~\ref{fig:setup}(c,d) and Fig.~\ref{Fig_2}.
By contrast, the full Hamiltonian \eqref{eq:H} does not obey this symmetry. 
The couplings $\Omega_{1}$ and $\Omega_2$ are symmetric and anti-symmetric, respectively, with the $a/2$ spatial shifts
\begin{equation}
\begin{aligned}
& \Omega_1(x+a/2,y) = \Omega_1(x,y+a/2) = \Omega_1(x,y) \, , \\
& \Omega_2(x+a/2,y) = \Omega_2(x,y+a/2) = - \Omega_2(x,y) \, .
\end{aligned}
\end{equation}
Thus the Hamiltonian $\hat{H}$ commutes with two combined shift operators
\begin{equation}
\hat{T}_{\mathbf{a}_l /2}=\hat{U} \exp\left(  - \frac{ i \, \mathbf{a}_l \cdot \hat{\mathbf{p}}}{2\hbar}   \right)\,,\quad\mathrm{where}\quad l=1,2\,, \label{eq:T_a_pm}
\end{equation}
and
\begin{align}
\hat{U} &= \left|2\right\rangle \left\langle 2\right|-\left|e\right\rangle \left\langle e\right|-\left|1\right\rangle \left\langle 1\right|, & {\rm with} & & \hat U^2 = \hat I\,.\label{eq:U}
\end{align}
The operator \eqref{eq:T_a_pm} combines a spatial translation by $\mathbf{a}_l / 2$ with a $\pi$ phase-flip of the states $\ket{e}$ and $\ket{1}$.
Thus the square of the combined operator $\hat{T}_{\mathbf{a}_l/2}^2 = e^{  - i \, \mathbf{a}_l \cdot \hat{\mathbf{p}} / \hbar }$ returns to a state-independent spatial shift by $a$. The Hamiltonian $\hat{H}$ and the combined shift operator $\hat{T}_{\mathbf{a}_l /2}$ therefore share a set of eigenstates following the Bloch ansatz
\begin{equation}
\ket{\psi_s^{(\boldsymbol{q})}(\boldsymbol{r})} =e^{i\boldsymbol{q}\cdot\boldsymbol{r}} \ket{g_s^{(\boldsymbol{q})}(\boldsymbol{r})} \,,\label{eq:|psi-q>}
\end{equation}
with
\begin{equation}
\hat{H} \left|\psi_s^{(\boldsymbol{q})}(\boldsymbol{r})\right\rangle = E_s(\boldsymbol{q}) \left|\psi_s^{(\boldsymbol{q})}(\boldsymbol{r})\right\rangle \, ,  \label{eq_eigen_H_Bloch}
\end{equation}
and 
\begin{equation}
\hat{T}_{\mathbf{a}_l /2} \left|\psi_s^{(\boldsymbol{q})}(\boldsymbol{r})\right\rangle = e^{i\boldsymbol{q}\cdot \mathbf{a}_l / 2}\left|\psi_s^{(\boldsymbol{q})}(\boldsymbol{r})\right\rangle \,,\label{eq:T_a_pm-eigenvalue}
\end{equation}
with eigenenergy $E_s(\boldsymbol{q})$, crystal momentum $\boldsymbol{q}$, and dark state band index $s=1,2,3,...$.
In what follows we focus on the lowest band with $s=1$ and therefore omit the band index~\footnote{In practice, we diagonalize the full Hamiltonian and identify the dark state manifold as those bands with the lowest minimum imaginary contribution to the energy.}.

The periodic part of the Bloch solution \eqref{eq:|psi-q>} satisfies
\begin{equation}
\hat{T}_{\mathbf{a}_l /2} \left|g^{(\boldsymbol{q})}(\boldsymbol{r})\right\rangle =\hat{U}\left|g^{(\boldsymbol{q})}(\boldsymbol{r}+\mathbf{a}_l / 2)\right\rangle =\left|g^{(\boldsymbol{q})}(\boldsymbol{r})\right\rangle \,\label{eq:U-g-q-eigenvalue}
\end{equation}
for a spatial shift of a half of the lattice constant.
Expanding in terms of atomic internal states gives
\begin{equation}
|g^{(\boldsymbol{q})}(\boldsymbol{r})\rangle=\sum_{j=e,1,2} g_{j}^{(\boldsymbol{q})}\left(\boldsymbol{r}\right)\ket{j}\,,\label{eq:|g^q>-expansion}
\end{equation}
subject to the conditions
\begin{equation}
g_{j}^{(\boldsymbol{q})}(\boldsymbol{r}+\mathbf{a}_l / 2)=-g_j^{(\boldsymbol{q})}(\boldsymbol{r}) \quad \mathrm{for}\,\,j=e,1 \, ,
\label{g_restriction_a2_1}
\end{equation}
and
\begin{equation}
g_{j}^{(\boldsymbol{q})}(\boldsymbol{r}+\mathbf{a}_l / 2)=g_{j}^{(\boldsymbol{q})}(\boldsymbol{r})
\quad \mathrm{for}\,\,j=2\,.
\label{g_restriction_a2_2}
\end{equation}
The Bloch ansatz given by Eq.~(\ref{eq:|psi-q>})
is characterized by a 2D crystal momentum $\boldsymbol{q} = q_x \mathbf{e}_x + q_y \mathbf{e}_y$
covering an extended Brillouin zone (BZ) with $q_{x,y} \in [-\kr,\kr)$, a four fold increase in area compared to the BZ of a square lattice with period $a$. 
Correspondingly, the area of the unit cell is reduced by a factor of four~\footnote{One could choose $\hat{U}=-\hat{U}$ in \eqref{eq:U} for the combined translation operator, as this too would satisfy $\hat{U}^{2}=\hat{I}$. However, the form given by \eqref{eq:U} unfolds the energy bands such that the band corresponding to the ground state in the dark state manifold typically has minimum energy and minimum losses at the center of the BZ with $\boldsymbol{q}=\mathbf{0}$, e.g., see Fig.~\ref{Fig_3}. Furthermore, there are in total four $\hat{U}$ operators one could choose from: two each for independent shifts along $\ex$ and $\ey$}.

We note that when $\chi = \pi/2$, the Hamiltonian supports an additional symmetry with respect to the spatial shifts by $\mathbf{a}_1/4 \pm \mathbf{a}_2/4$, as can be seen in Fig.~\ref{Fig_2}(c,f).
In this particular case the BZ can be further unfolded into a rhombus.

For $\chi=0$ (and neglecting the decay rate $\Gamma$), the Hamiltonian $\hat{H}$ obeys time reversal symmetry.
In that case a simultaneous complex
conjugation and inversion of the quasi-momentum leaves the eigenvalue equation (\ref{eq_eigen_H_Bloch}) unchanged, giving 
\begin{equation}
E^{(-\boldsymbol{q})}=E^{(\boldsymbol{q})}\quad\mathrm{and}\quad g_{j}^{(-\boldsymbol{q})}(\boldsymbol{r})=\left[g_{j}^{(\boldsymbol{q})}(\boldsymbol{r})\right]^{*}\,.\label{eq:Time-reversal}
\end{equation}
We numerically confirmed that this condition is well maintained for atomic dynamics in the dark state manifold where atomic decay is suppressed.
Note that even for $\chi\ne0$ the condition $E^{(-\boldsymbol{q})}=E^{(\boldsymbol{q})}$ holds because complex conjugation does not change $|\Omega_{2}|$ and thus does not alter the energy spectra plotted in Figs.~\ref{Fig_3},~\ref{Fig_comparison}.

\section{Numerical results}\label{sec:results}

We now describe our numerical method for obtaining the energy spectra and present our findings~
\footnote{See Supplementary Material at \url{http://web.vu.lt/ff/g.juzeliunas/2D_Lambda_Supplementary.zip} for the Python3 codes used to obtain the results featured in this publication. The material contains instructions on installing and operating the software.}.

\subsection{Numerical method}
\label{Sec_numerical_method}

The band structure is most easily solved by factoring out the plane wave component $e^{i \boldsymbol{q} \cdot \boldsymbol{r}}$ in the Bloch eigen-function \eqref{eq:|psi-q>}
\begin{equation}
\hat{H}^{(\boldsymbol{q})} \left|g^{(\boldsymbol{q})}(\boldsymbol{r})\right\rangle = E^{(\boldsymbol{q})} \left|g^{(\boldsymbol{q})}(\boldsymbol{r})\right\rangle \, , \label{eq_eigen_H_Bloch_periodic}
\end{equation}
transforming the Hamiltonian from Eq.~\eqref{eq_eigen_H_Bloch} to
\begin{equation}
\label{eq_H_q}
\hat{H}^{(\boldsymbol{q})} (\boldsymbol{r})  = \frac{1}{2M}  \left( -i \hbar \boldsymbol{\nabla} + \hbar \boldsymbol{q} \right)^{2} +  \hat{V}(\boldsymbol{r})
 \, .
\end{equation}
We obtain the energy spectrum numerically using the Fourier representation of the eigen-value equation \eqref{eq_eigen_H_Bloch_periodic} and of the periodic Bloch functions
\begin{equation}\label{eq:Fourier_decompose_2D}
\ket{g^{(\boldsymbol{q})}(\boldsymbol{r})} = \sum_{n_x,n_y=-N}^{N} \: e^{i k_R (x n_x + y n_y)} \ket{g^{(\boldsymbol{q})} (n_x, \, n_y)} \, ,
\end{equation}
with $(2N+1)^2$ Fourier components.
The resulting right-handed matrix eigenvalue problem
\begin{equation}\label{eq:diagonalize_2D}
E^{(\boldsymbol{q})} \,  g^{(\boldsymbol{q})}_{n_x n_y j} = \left[ H^{(\boldsymbol{q})}  \right]^{n^{\prime}_x n^{\prime}_y j'}_{n_x n_y j} \, g^{(\boldsymbol{q})}_{n_x n_y j} \,
\end{equation}
is encoded with the combined set of indices $(n_x, n_y, j)$ including both the Fourier components ($n_x$, $n_y$) and the atomic internal states $j$.
The Hamiltonian-matrix $[H^{(\boldsymbol{q})}]$ is sparsely populated with $3^2(2N+1)^{4}$ elements and a typical filling ratio $\lesssim  10^{-5}$ for $N \approx 100$.
We use shift inversion to amplify solutions near the bottom of the dark state manifold using libraries optimized for sparse matrix diagonalization~\cite{ARPACK, scipy}.
Section~\ref{sec:symmetries} implies that certain Fourier components of the periodic Bloch function $g_j^{(\boldsymbol{q})} (n_x, \, n_y)$ must be zero to unfold the BZ; we strictly enforce this condition by zeroing out some of matrix elements as described in Appendix~\ref{Appendix_A}.

Diagonalization of the adiabatic dark state Hamiltonian \eqref{eq:Schroed_D} is much more challenging numerically due to many non-zero Fourier components associated with the effective potentials $\Udd$ and $\Add$. Although the adiabatic dark state Hamiltonian-matrix $[H^{(\boldsymbol{q})}_D]$ is 9 times smaller with $(2N+1)^{4}$ elements, for $N \approx 100$ it has a filling ratio of $\lesssim 0.04$, making it significantly more dense.
By removing Fourier components with negligible amplitudes, we reduce the filling ratio to $\lesssim 0.01$.

\subsection{Energy dispersions} \label{sec_energy_dispersions}

\begin{figure}[t]
\centering
  \includegraphics[scale=1,keepaspectratio]{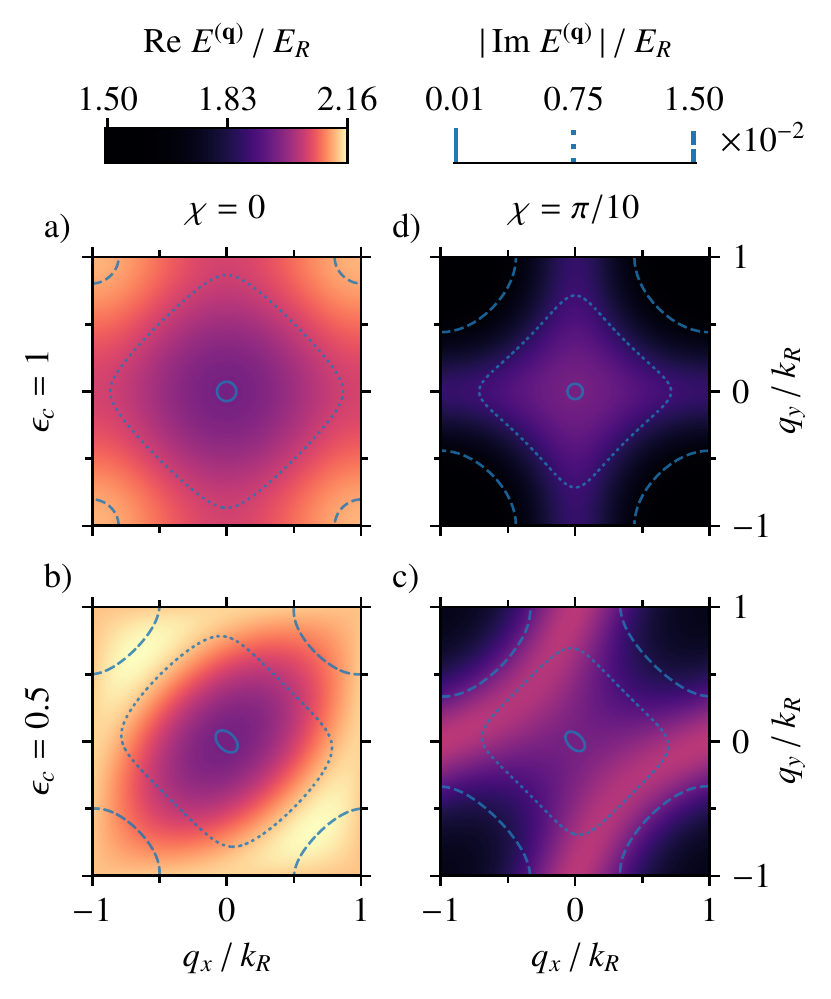}
  \caption{Real (Hermitian) and imaginary (anti-Hermitian) parts of the energy dispersion (represented by a non-linear colormap and blue contour lines, respectively) describing the lowest dark Bloch band. The parameters for (a,b,c,d) match that of Figs.~\ref{Fig_2}(a,d,e,b).
  Other parameters
  are $\epsilon=0.1$, $\Delta=0$, $\hbar \Omega_p = 2000 \, \Er$, $\hbar \Gamma = 1000 \, \Er$ and $N=250$.
  The maximum absolute value of the anti-Hermitian part of the energy is $ \approx 0.022 \Er$ in part c) at the corners of the BZ; the minimum is $\approx 8.6 \cdot 10^{-6} \Er$ in part b) for $\boldsymbol{q}=\mathbf{0}$; the average over parts a)--d) is $\approx 0.01 \Er$.}
  \label{Fig_3}
\end{figure}

The real and imaginary parts of the energy dispersion describing the lowest Bloch band in the dark manifold are plotted in Fig.~\ref{Fig_3} for four combinations of $\epsilon_c$ and $\chi$ (we avoid $\chi=\pi/2$, corresponding to Figs.~\ref{Fig_2}(c,f), which results in a gapless energy dispersion).
The real part is qualitatively different for each combination of parameters in Fig.~\ref{Fig_3}: in (a) the dispersion is reminiscent of that of a 2D square lattice; in (b) the curvature near ${\boldsymbol{q}=\mathbf{0}}$ has become anisotropic and tiny local minima have appeared at the corners of the BZ; in (c) the dispersion at ${\boldsymbol{q}=\mathbf{0}}$ has become a saddle point and the energies at the corners of the BZ continue to fall; the trend is completed in (d) where the dispersion has a global maximum at ${\boldsymbol{q}=\mathbf{0}}$.

The imaginary part (blue contours) in Fig.~\ref{Fig_3} results from a small admixture of the excited state and is everywhere negative; it contains no contribution from the gauge field $\Add$, and thus quantifies only anti-Hermitian losses.
From the perspective of the dark state adiabatic potentials, this admixture results from non-adiabatic coupling to the bright states.
However, despite the large value of $\hbar \Gamma/\Er = 1000$ used in Fig.~\ref{Fig_3}, we observe a relatively small population transfer into the excited state: for our parameters the imaginary energy can be as low as $8.6 \times 10^{-6} \Er$ at the center of the BZ [Fig.~\ref{Fig_3}(b)].
For a wide range of $\epsilon_c$ and $\chi$, the imaginary part of the energy averaged over the BZ is $\approx - 0.01 \Er$.
The excited state occupation probability (and therefore losses) is further reduced at blue-detuning $\Delta > 0$, by reducing sharpness in the potential peaks with a larger value of $\epsilon$, or by increasing the $\Omega$'s, as observed for 1D dark state lattices~\cite{Zoller2016, Wang2018, Jendrzejewski2016, Zubairy2020, Gvozdiovas2021, Kubala2021}. 

\begin{figure}[t]
\centering
  \includegraphics[scale=1,keepaspectratio]{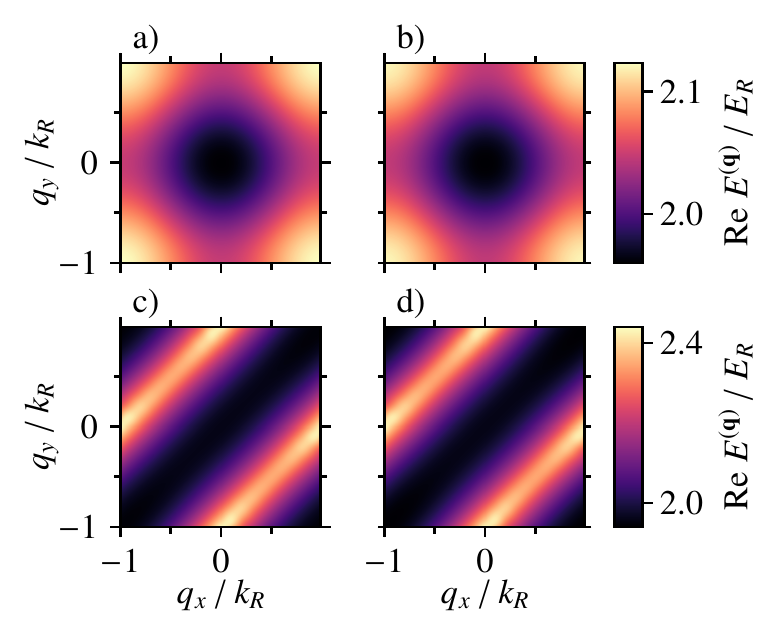}
  \caption{Real part of the energy dispersions from the full Hamiltonian [Eq.~\eqref{eq:H}] (a,c), and adiabatic dark state Hamiltonian [Eq.~\eqref{eq:Schroed_D}] (b,d). 
  (a,b) were computed for $\chi=0$, $\epsilon_c=1$; (c,d) for $\chi=1.4$ rad and $\epsilon_c=0.09$. 
  The other parameters are $N=100$, $\epsilon=0.1$; for (a,c), $\hbar \Omega_p=2000 \Er$, $\hbar \Gamma = 1000 E_R$, and $\Delta=0$.}
  \label{Fig_comparison}
\end{figure}

Figure~\ref{Fig_comparison} compares the real part of the ground dark band dispersion computed using the full Hamiltonian [Eq.~\eqref{eq:H}] in (a,c), with that computed using the adiabatic approximation [Eq.~\eqref{eq:Schroed_D}] in (b,d).
In each case we computed the band structure for $\chi=0$, $\epsilon_c=1$ and $\chi=1.4\ {\rm rad}$, $\epsilon_c=0.09$ to highlight regimes where the adiabatic approximation is at its best [in (a,b)] and worst  [in (c,d)], respectively.
We find that these dispersions are visually indistinguishable even in the presence of needle-like barriers [Fig.~\ref{Fig_2}(c)].
Quantitatively, the largest discrepancy is at the edge of the BZ, where losses (absent in the adiabatic approximation) are maximal.
Conversely, states with crystal momentum $\boldsymbol{q} \approx \mathbf{0}$ retain near-perfect adiabaticity even in the worst case scenario [Fig.~\ref{Fig_comparison}(c,d)].
The regions of validity of the adiabatic approximation are analogous to those in 1D systems~\cite{Wang2018, Jendrzejewski2016, Zubairy2020, Gvozdiovas2021, Kubala2021}: $\Delta=0$; large $\Omega$'s, non-infinitesimal $\epsilon$, and small, but non-zero $\Gamma$ to avoid bright-dark resonances \cite{Zoller2016}, all tend to reduce leakage from the dark state.
The influence on adiabaticity of $\chi$ and $\epsilon_c$ is non-trivial and the parameters used in (c) reflect the global maximum in the $\chi$ and $\epsilon_c$ parameter space.

Generally, the relationship between losses and the parameters $\epsilon$, $\chi$ and $\epsilon_c$ is complicated: all of these parameters have a significant influence on the non-adiabatic corrections.
Overall, losses averaged over the BZ are minimal when $\epsilon_c = 1$ and $\chi = 0$; in this limit the effective magnetic field vanishes, leaving only the scalar potential which takes the form of an array of 2D cages with gaps as shown in Fig.~\ref{Fig_2}(a) with the resulting energy dispersion in Fig.~\ref{Fig_3}(a) and Fig.~\ref{Fig_comparison}(a,b). This scenario also yields a highly flat ground band with a large band gap ($\approx 3 \Er$ with $\epsilon \ll 1$).
In this limit, the dark state Hamiltonian \eqref{eq:Schroed_D} can effectively be approximated as a sum of two orthogonal 1D potentials with sub-wavelength barriers:
\begin{equation}
\hat{H}_D(\boldsymbol{r}) \approx \hat{H}^{\rm (sep)}_D(x)+\hat{H}_D^{\rm (sep)}(y)\,,
\label{eq_HD_sep}
\end{equation}
giving energy bands
\begin{equation}
E(\boldsymbol{q})=E^{\rm (sep)}(q_{x})+E^{\rm (sep)}(q_{y})\,.\label{sum_rule}
\end{equation}
The 1D Hamiltonian
\begin{equation}
\hat{H}_D^{\rm (sep)}(x) = \frac{\hat{p}_x^{2}}{2M}+U_D^{\rm (sep)}(x)
\end{equation}
contains a Kronig-Penney like potential
\begin{equation}
U^{(\text{sep})}_D(x)=\frac{ \Er \epsilon_{\text{sep}}^{2}\,\cos^{2}{(\kr x)}}{\left[\epsilon_{\text{sep}}^{2}+\sin^{2}{(\kr x)}\right]^{2}} \, ,
\label{U_D_2D_s}
\end{equation}
which appears in 1D analogues of our setup~\cite{Zoller2016, Wang2018}, with $\Omega_2(x) = \Omega_c \sin{(\kr x)}$ and $\epsilon_{\text{sep}} = \Omega_p / \Omega_c$.
The real part of the energy scales approximately as $s^2_x + s^2_y$, with positive integers $s_x$ and $s_y$.
For a sufficiently deep lattice with $\epsilon \lesssim 0.2$, we numerically confirmed that the separable \eqref{eq_HD_sep} and non-separable \eqref{eq:Schroed_D} Hamiltonians give similar energy dispersions when $\epsilon \approx \epsilon_{\text{sep}}$.
Such an effective scenario depicts an array of gapless 2D cages with a constant wall height equal to half of the maximum height of the true scalar potential \eqref{eq:A_D}.
Finally, we note that the full Hamiltonian \eqref{eq:H} cannot be treated this way due to its internal structure.

\subsection{Non-Hermitian tight binding model}

\begin{figure}[t]
\centering
  \includegraphics[scale=1,keepaspectratio]{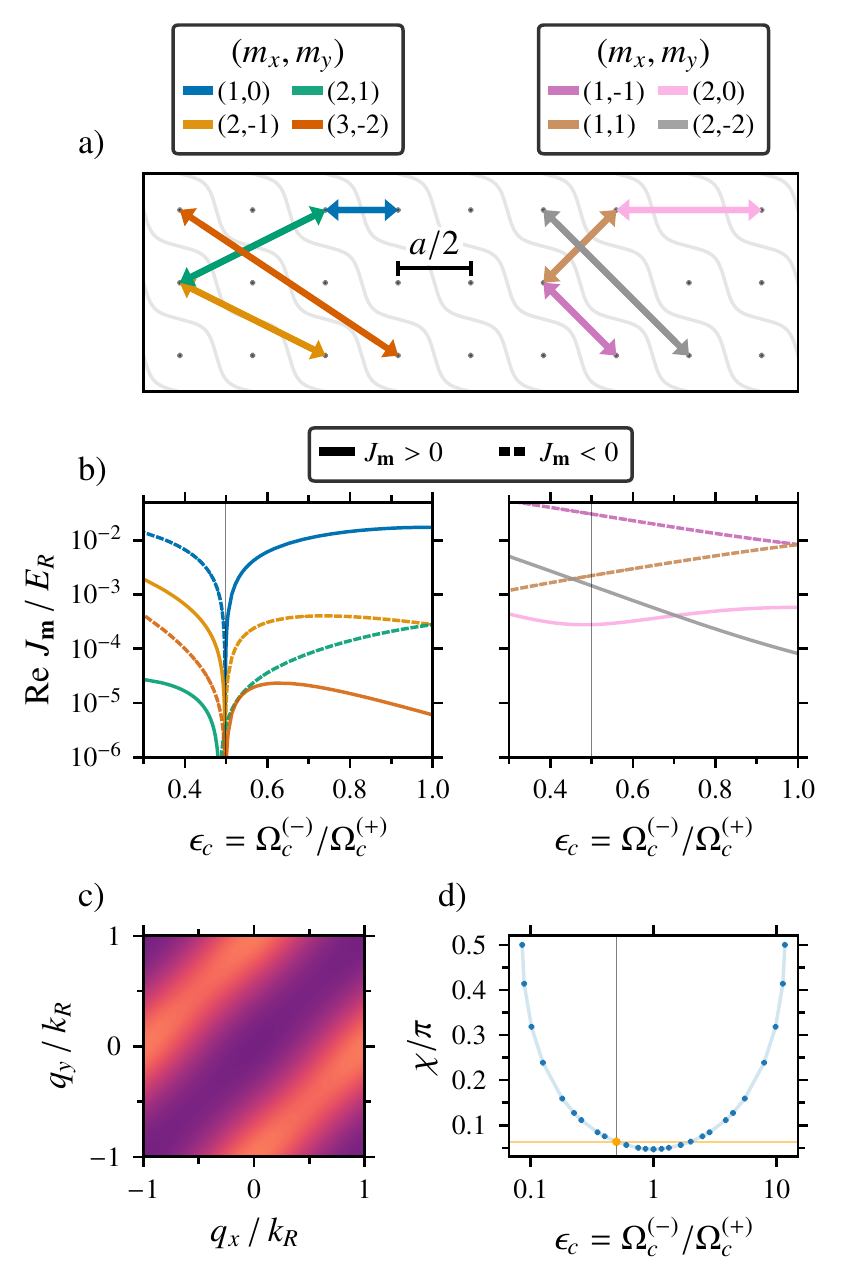}
  \caption{a), Hopping directions in a periodic array of tubes. The left(right) half illustrates odd(even) order tunneling between lattice sites.
  b), Hermitian (real) part of hopping amplitudes $J_{m_x, m_y}$ corresponding to the directions in part a) and describing the ground state in the dark manifold vs $\epsilon_c$ with $\chi = 0.2$ rad. Dashed(solid) lines mark negative(positive) values. 
  c), Real energy dispersion for the thin vertical line in b) with energy represented by the colormap in Fig.~\ref{Fig_3}.
  d), Combinations of $\epsilon_c$ and $\chi$ for which the real part of $J_{0,1}$ vanishes. The orange point coincides with c) and the orange line marks $\chi = 0.2$ rad used in b)--c). Other parameters for b)--d) are $\epsilon = 0.1$, $\Delta=0$, $\hbar \Omega_p = 2000 \, \Er$, $\hbar \Gamma = 1000 \, \Er$ and $N=260$.
  }
  \label{Fig_4}
\end{figure}

We express the band structure, such as shown in Fig.~\ref{Fig_3}, as a Fourier transform~\cite{Goringe1997, Delerue2001, Bellec2013}:
\begin{equation}
E^{(\boldsymbol{q})} = \sum_{m_x=-\infty}^{\infty} \, \sum_{m_y=-\infty}^{\infty} J_{m_x, \, m_y}\exp(-i \boldsymbol{q} \, \cdot \, \delta \mathbf{R}_{m_x, \, m_y}) \, ,
\label{eq:E_k_2D}
\end{equation}
where $J_{\mathbf{m}}$ describing hopping with range ${\mathbf{m} \equiv (m_x, m_y)}$ has both real and imaginary parts.
The element $J_{0,0}$ is the on-site energy.
These complex tight binding parameters describe conventional tunneling, and as is well established in photonic systems, can incorporate both gain and loss~\cite{Rudner2009, Schomerus2013, El-Ganainy2018, Lieu2018, Ashida2020, Han2021}.
We obtain the tight binding parameters from the Fourier transform
\begin{equation}
J_{m_x , \, m_y}= \kr^{-2} \int_{\rm BZ} E^{(\boldsymbol{q})} \exp(i \boldsymbol{q} \, \cdot \, \delta \textbf{R}_{m_x , \, m_y}) \, \mathrm{d} \boldsymbol{q} \, \label{eq:t_n_2D}
\end{equation}
of the numerically obtained band structure $E^{(\boldsymbol{q})}$. For any value of $\chi$ and $\epsilon_c$, we find that $ \text{Re} \, J_{m_x , \, m_y}$ fully describes the band structure of the Hermitian part of the Hamiltonian, and therefore $\text{Im} \, J_{m_x , \, m_y}$ fully accounts for the anti-Hermitian contribution of $i \Gamma/2$.

We begin by commenting on the impact of the available physical parameters $\epsilon$, $\epsilon_c$, $\chi$.
Starting with the square lattice scenario depicted in Fig.~\ref{Fig_2}(a), such that $\epsilon_c = 1$ and $\chi=0$; $\epsilon$ determines the lattice depth, simultaneously modifying all of the hopping parameters while moving between shallow ($\epsilon \rightarrow 1$) and deep lattice ($\epsilon \rightarrow 0$) regimes.
Next, tuning $\chi \rightarrow \pi/2$ delocalizes atoms as the scalar potential walls shrink to point-like barriers as shown in Fig.~\ref{Fig_2}(c); this can be counteracted by moving away from $\epsilon_c = 1$ towards $\epsilon_c \rightarrow 0$ or $\epsilon_c \rightarrow \infty$, restoring the longitudinal extent of barriers as they approach 1D walls, as can be seen by comparing Figs.~\ref{Fig_2}(b,e). 
Alternatively, the spatial extent of barriers can be restored by increasing $\epsilon$ slightly, consequently reducing their sharpness.

The dependence of the real part of the hopping parameters on $\epsilon_c$ is shown in Fig.~\ref{Fig_4}(b) (with $\epsilon = 0.1$ and $\chi=0.2$), and the corresponding imaginary part is plotted in Fig.~\ref{Fig_losses}.
Many of the hopping parameters are identical due to the symmetries discussed in Sec.~\ref{sec:symmetries}---leading to $J_{m_x, m_y} = J_{m_y, m_x} = J_{-m_x, -m_y} = J_{-m_y, -m_x}$---and are thus omitted (this includes the non-Hermitian part).
Furthermore, when $\epsilon_c = 1$ the lattice becomes symmetrical with respect to $x$ and $y$ leading to $J_{\pm m_x, \pm m_y} = J_{\pm m_y, \pm m_x} $.

\subsection{Band flattening}

Figure~\ref{Fig_4}(b) shows our main finding: in a narrow region of $\epsilon_c$ ($\epsilon_c \approx 0.498$, thin vertical line) the real part of the nearest-neighbor (NN) hopping $J_{1,0}$ vanishes, and the remaining odd hopping terms such as $J_{2,-1}$, $J_{2,1}$, $J_{3,-2}$ approach zero.
The remaining even-order tunneling processes become dominant, with a leading contribution from $J_{1,-1}=J_{-1,1}$ describing diagonally oriented tunneling within tubes (magenta); the next leading contributions are $J_{1,1}$ and $J_{2,-2}$, giving coupling between next nearest neighboring tubes, and longer range tunneling within tubes, respectively. 
This effectively describes an array of nearly decoupled tubes represented by the highly anisotropic energy dispersion in Fig.~\ref{Fig_4}(c).
The odd terms $J_{2,1}$ and $J_{2,-1}$ couple neighboring tubes, but are weaker than $J_{1,-1}$ by up to 4 orders of magnitude.
As we describe in Sec.~\ref{sec_tube}, this results from Aharonov-Bohm like quantum interference from the geometric vector potential.

Similarly, one can observe the decoupling point by fixing $\epsilon_c$ and tuning $\chi$.
In fact, for every $\chi \gtrsim 0.05 \pi$, two values of $\epsilon_c$ give the decoupled tube scenario [Fig.~\ref{Fig_4}(d)]; these are related by $\epsilon_c \rightarrow 1/\epsilon_c$ and result from identical lattices rotated by 90 degrees.
These two branches merge at $\epsilon_c = 1$ where the lattice is symmetric with respect to 90 degree rotations.
A more detailed discussion on these issues is presented in Sec.~\ref{sec_tube}.

\begin{figure}[t]
\centering
\includegraphics[scale=1,keepaspectratio]{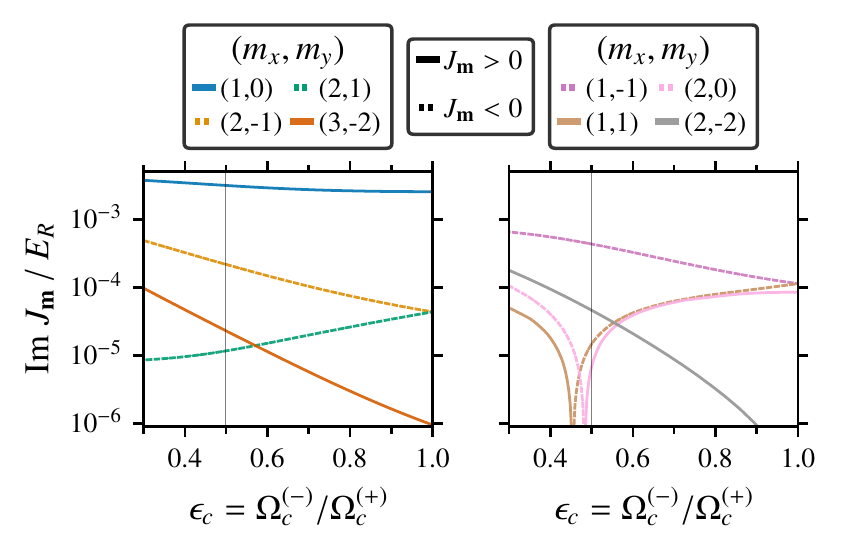}
  \caption{Anti-Hermitian contribution to the tight binding parameters $J_{m_x, m_y}$ shown in Fig.~\ref{Fig_4}(a,b). Dashed(solid) lines mark negative(positive) values.}
  \label{Fig_losses}
\end{figure}

We also examined the energy band gap at the special points.
For $\chi = 0.2$ rad, $\epsilon = 0.1$ and $\epsilon_c \approx 0.498$, both the indirect and direct energy gaps are $\approx 1.25 \Er$.
They can be made wider (with an upper limit of $3 \Er$) while maintaining the weakly coupled tube scenario by approaching $\epsilon_c \rightarrow 1$, and by reducing $\chi$ and $\epsilon$.

We now turn to the non-Hermitian part of the energy where our findings are no less interesting.
The imaginary contribution to the on-site energy $ \text{Im} \, J_{0,0} \approx - 0.01 \Er$ is always negative, describing on-site atom loss.
However, the imaginary part of the energy is nearly zero for crystal momentum $\boldsymbol{q}=\mathbf{0}$, implying that the sum of the imaginary tight binding parameters $J_{m_x, \, m_y}$ entering Eq.~\eqref{eq:E_k_2D} is nearly zero.
In Fig.~\ref{Fig_losses} we demonstrate that this results from hopping matrix elements with imaginary components of both signs -- an example of tight-binding gain-loss balance also observed in 1D dark state lattices with decay~\cite{Gvozdiovas2021}, which appears despite a strictly lossy Hamiltonian \eqref{eq:H}.

\subsection{Quantum interference}
\label{sec_tube}

\begin{figure}[t]
\centering
  \includegraphics[scale=1,keepaspectratio]{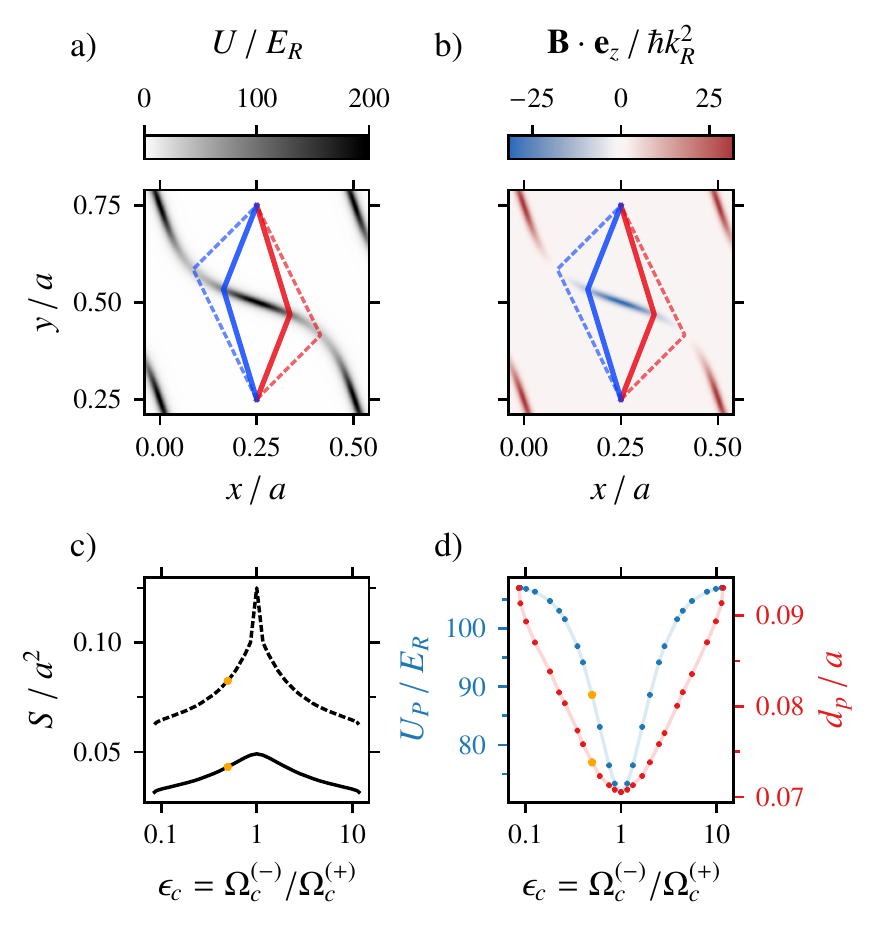}
  \caption{
  a), b), Branching paths (blue, red) representing the nearest-neighbor tunneling $J_{0,1}$ drawn on top of the effective potentials $\Udd$ and $\Bdd$ for parameters matching Fig.~\ref{Fig_4}(c).
  c), Areas of solid and dashed parallelograms in a)--b). 
  d), Barrier height $U_p$ (blue), and width $d_p$ (red), both for the solid contour. c)--d) are plotted for the special points in Fig.~\ref{Fig_4}(d). The orange points mark parameters describing a)--b).
  }
  \label{Fig_5}
\end{figure}

Here we qualitatively explain the suppression of intertube tunneling by considering trajectories linking neighboring tubes as sketched in Fig.~\ref{Fig_5}(a,b).
A quantitatively complete path integral description involving the sum over all paths is not needed to understand the basic origin of the suppression.

To this end, we consider simple ray-like paths connecting the centers of neighboring lattice sites that undergo Snell's law type refraction at the potential barriers (for this argument we do not consider the reduction in transmission amplitude due to reflections).
We compute the phase difference
\begin{equation}
\phi_B = \frac{1}{\hbar} \left( \int_{C_1} \Add \cdot \mathrm{d} \boldsymbol{r} - \int_{C_2} \Add \cdot \mathrm{d} \boldsymbol{r} \right) = \frac{1}{\hbar} \oint_{C} \Add \, \mathrm{d} \boldsymbol{r}  \, \label{phi_B_def}
\end{equation}
associated with paths $C_1$ (red) and $C_2$ (blue) that combine to encircle the tall barrier.
As illustrated in Fig.~\ref{Fig_5}(a,b), these together form a closed contour $C$.
The accumulated phase $\phi_B$ is thus the line integral of $\Add$ along $C$ (equal to the integral of $\Bdd$ within $C$ by Stokes' theorem); when $\phi_B = \pi + 2\pi n$, for integer $n$, these two paths destructively interfere, suppressing tunneling.

We investigated two families of contours.
\begin{enumerate}

\item[A] Dashed: these cross the scalar potential barrier at its minimum [Fig.~\ref{Fig_5}(a)] and fully enclose the magnetic field peak [Fig.~\ref{Fig_5}(b)].

\item[B] Solid: these paths are derived from the dashed contour by symmetrically moving the left and right corners along the scalar potential wall until $\phi_B = \pi$.
\end{enumerate}

Contour (A) was selected to minimize the potential energy cost of the path at the expense of increased length and kinetic energy owing to the larger corner angle.
For the parameters used in Fig.~\ref{Fig_4}(c), this contour results in $\phi_B \approx 1.3 \pi$: larger than needed for destructive interference.
Indeed, Fig.~\ref{Fig_5}(c) plots the area enclosed by these trajectories as a function of $\epsilon_c$, and shows that the optimal trajectory (B) is always reduced in size.
This indicates that the representative (i.e. saddle point) trajectory minimizes a combination of potential and kinetic energy.
Furthermore, at $\epsilon_c=1$ (the point where the lattice has 90 degree rotational symmetry) the trajectories' areas are maximized, and as shown in Fig.~\ref{Fig_5}(d) the barrier height (blue) and width (red) pertaining to contour (B) are minimized.
This corresponds to paths with the most extreme trade-off: minimal potential energy and maximal kinetic. 

Our argument qualitatively describes first order tunneling such as $J_{1,0}$.
More generally, this description also explains suppression of only odd-order tunneling processes. 
As an example, consider the even-order hopping parameters $J_{1,1}$ and $J_{2,0}$.
$J_{1,1}$ tunneling is achieved by a single classical path that cuts through the scalar potential minima (thus without any option for interference effects).
We can explain $J_{2,0}$ in terms of a stacked pair of solid trajectories, but in this case $\phi_B = 2\pi$, leading to constructive interference.
In general, even order tunneling terms are associated with even integer multiples of $\pi$ (either constructive interference, or none at all) and odd-order trajectories have odd-integer multiples of $\pi$ (destructive interference).
Lastly, $J_{1,-1}$ dominates because its path is completely unobstructed.

Moreover, we observe destructive interference of NN tunneling in the higher energy bands of the dark manifold for similar parameter values, supporting the generic applicability of our classical ray model.



\section{Discussion and Outlook}\label{sec:outlook}

The 2D lattice featuring sub-wavelength structures considered here yields highly tunable geometric scalar and vector potentials with minimal spontaneous emission. 
The scalar potential can yield: a 2D square lattice with sub-wavelength barriers, an array of Delta function-like peaks (a 2D Dirac comb), or a lattice of interacting zigzag tubes.
Furthermore, the band structure is greatly affected by the geometric vector potential where tunneling between tubes can be suppressed due to Aharonov-Bohm type destructive interference. 

These lattices can be used to realize novel many body phases.
When tunneling is suppressed in conventional deep lattices, the associated maximally localized Wannier orbitals are very strongly confined to individual lattice sites.
In the present case, both intra- and inter-tube interactions are enhanced even at near-zero inter-tube tunneling, owing to the relatively shallow barriers and concomitantly extended Wannier orbitals.

From a broader perspective, this technique can create lattices with features well below the optical diffraction limit wherever the interfering laser beams in the $\ex$-$\ey$ plane approach zero.
Changing the number and intersection angles of these in-plane beams therefore allows for a range of lattice geometries, including quasi-crystalline.
In addition, the dark state lattice discussed here can also be extended to disordered configurations by using an optical speckle field for $\Omega_2$ in Fig.~\ref{fig:setup} rather than a standing wave potential.
The resulting dark states feature disordered geometric potentials, including a disordered magnetic field.
From our observation that the synthetic magnetic field can be used to destroy many of the hopping parameters, one can expect that a disordered magnetic field could create non-trivial tunneling paths.
In the broader context of localization in 2D disordered systems~\cite{Abrahams1979}, the localization properties of such a time-reversal symmetry breaking disorder potential is unclear~\cite{Galitski2005}.

\begin{acknowledgments}

The authors thank E.~Gutierrez, S.~Subhankar and E.~Benck for carefully reading the manuscript.
This work was supported by the Research Council of Lithuania (Grant No. S-MIP-20-36). 
IBS acknowledges support by the National Institute of Standards and Technology, and the National Science Foundation through the Quantum Leap Challenge Institute for Robust Quantum Simulation (Grant No. OMA-2120757).

GJ and IBS conceptualized the work; EG carried out all numerical simulations and analytical derivations, and created all the figures.
All authors contributed equally to writing the manuscript.

\end{acknowledgments}

\begin{appendix}

\section{Unfolding the BZ}

\label{Appendix_A}

Here we explain our numerical recipe for obtaining the Hamiltonian-matrix $[H^{(\boldsymbol{q})}]$ that describes the unfolded BZ. Using the symmetries discussed in Sec. \ref{sec:symmetries}, we modify the Hamiltonian-matrix by zeroing out some of the matrix elements. 

The conditions for non-zero matrix elements are determined from Eqs.~\eqref{eq:U}, \eqref{g_restriction_a2_1} and \eqref{g_restriction_a2_2}.
We first define $a/2$ symmetry constants $\mathcal{M}_j$ for each internal state. After choosing $\hat{U}$ according to  Eq.~\eqref{eq:U}, $\mathcal{M}_j$ are given by:
\begin{equation}
\label{M_j}
\mathcal{M}_j = \begin{cases}
0 &\text{for $j=2$}\\
1 &\text{for $j=e,1$}
\end{cases} \, .
\end{equation}
Thus $\mathcal{M}_j$ is even for $j=2$ and odd for $j=e,1$ following Eqs.~\eqref{g_restriction_a2_1}--\eqref{g_restriction_a2_2}. It then follows, for example in the case of $j=2$, that the even Fourier components $g^{(\boldsymbol{q})}_2(2 n_x, 2 n_y)$ describing $g^{(\boldsymbol{q})}_2(\boldsymbol{r})$ must be non-zero since it is an even function with regards to shifts by $a/2$ in $\ex$ and $\ey$, giving $\mathcal{M}_2=0$. The opposite is true for $j=e,1$. This can be written mathematically as
\begin{gather} \label{g_nonzero}
\begin{aligned}
& g^{(\boldsymbol{q})}_j(n_x, n_y) \ne 0 \, , \quad \text{if} \\
& (-1)^{\mathcal{M}_j + n_x}= 1 \quad \text{and} \quad (-1)^{\mathcal{M}_j + n_y}=1 \, .
\end{aligned}
\end{gather}
The non-zero Hamiltonian-matrix elements are then
\begin{gather} \label{H_nonzero}
\begin{aligned}
& \left[ H^{(\boldsymbol{q})}  \right]^{n^{\prime}_x n^{\prime}_y j'}_{n_x n_y j} \ne 0 \, , \quad \text{if} \\
& (-1)^{\mathcal{M}_{j} + n_x}= 1 \quad \text{and} \quad (-1)^{\mathcal{M}_j + n_y}=1 \quad \text{and} \\
& (-1)^{\mathcal{M}_{j^{\prime}} + n^{\prime}_x}= 1 \quad \text{and} \quad (-1)^{\mathcal{M}_{j^{\prime}} + n^{\prime}_y}=1 \, .
\end{aligned}
\end{gather}
We thus arrive at a new Hamiltonian-matrix with eigensolutions characterized by the extended BZ $q_{x,y} \in [-\kr,\kr)$. The Fourier-space eigenvectors \eqref{eq:Fourier_decompose_2D} diagonalizing this matrix are not truncated: all of the $3(2N+1)^2$ Fourier components (including the ones equal to zero) describe the solution, and so the numerically obtained eigenvectors for each $\boldsymbol{q}$-value describe a real space unit cell of area $a^2$ (as though the BZ were not unfolded). We note that this method of zeroing out matrix elements is sub-optimal -- a better solution would be to truncate the Fourier space, reducing the size of the Hamiltonian-matrix.

The same filtering procedure is valid for the adiabatic Hamiltonian-matrix $[H^{(\boldsymbol{q})}_D]$. It is described by one internal state $j=j_D$ -- the dark state \eqref{eq:D} -- which is invariant with respect to the combined shift operator $\hat{T}_{\mathbf{a}_l /2}$ defined in \eqref{eq:T_a_pm}:
\begin{equation}
 \hat{T}_{\mathbf{a}_l /2} \left|D(\boldsymbol{r})\right\rangle =\left|D(\boldsymbol{r})\right\rangle \,  , \label{eq:shift-dark}
 \end{equation}
therefore one has $\mathcal{M}_D = 0$ ($g_D^{(\boldsymbol{q})}$ is an even function, see also Fig.~\ref{fig:setup}(c,d)). 

\end{appendix}

\bibliography{main}

\end{document}